\documentclass[pdflatex,sn-aps]{sn-jnl}

\usepackage{graphicx}%
\usepackage{multirow}%
\usepackage{amsmath,amssymb,amsfonts}%
\usepackage{amsthm}%
\usepackage{mathrsfs}%
\usepackage[title]{appendix}%
\usepackage{xcolor}%
\usepackage{textcomp}%
\usepackage{manyfoot}%
\usepackage{booktabs}%
\usepackage{algorithm}%
\usepackage{algorithmicx}%
\usepackage{algpseudocode}%
\usepackage{listings}%

\theoremstyle{thmstyleone}%
\theoremstyle{thmstyletwo}%

\theoremstyle{thmstylethree}%

\begin{document}

\title[Normalizing Fock space states in static spacetimes]{Normalizing Fock space states in static spacetimes}


\author*[1]{\fnm{Jesse} \sur{Huhtala}}\email{jejohuh@utu.fi}
\author[1]{\fnm{Iiro} \sur{Vilja}}\email{vilja@utu.fi}

\affil[1]{\orgdiv{Department of Physics and Astronomy}, \orgname{University of Turku}, \orgaddress{\city{20014 Turku},  \country{Finland}}}


\abstract{In quantum field theory, sharp momentum states have to be normalized to be in Fock space. We investigate different normalization schemes, both box normalization and wave packets. These methods are equivalent in flat spacetimes, but turn out to produce different results in curved spacetimes, specifically in those that break translation invariance. This means that scattering processes have to be defined in relation to the normalization scheme used, rather than being independent of it as is the case in flat spacetime. We provide an illustrative example of this phenomenon.}


\maketitle
\tableofcontents
\section{Introduction}\label{sec:intro}
Curved spacetime quantum field theory is by now a mature field with multiple textbooks \cite{Mukhanov2007, Birrell1982} and a research community of healthy size. Despite many decades of research, there are still foundational questions of interest relating to measurement \cite{Fewster2020} and the appropriate mathematical framework in which to place quantum field theory \cite{Fewster2021,Hollands2001,Hollands2002,Hollands2010}. Concepts that are by now straightforward in flat spacetime are not always so in curved spacetimes, and it is one of these foundational concepts we wish to investigate in this article.

Our aim is to deal with the normalization of Fock space states in static spacetimes in the conventional QFT framework. As is well known, sharp momentum states, while heuristically convenient, are not actually in the Fock space. To obtain proper Fock space states, one should either use wave packets \cite{Peskin1995} or "put the system in a box" \cite{Folland2008}, typically with periodic boundary conditions. In flat spacetime, it does not matter which method one chooses; the end result is entirely equivalent after all the appropriate limits are taken. No wave packet or box dependence remains. We will show that this is not the case in curved spacetimes for scattering processes -- rather, the end result depends on which method one chooses for normalization.

This matters because it has been shown that wave packet corrections may change the results of scattering calculations even in flat spacetimes. Ishikawa and collaborators showed that wave packet interference terms change decay rates in flat spacetimes for light particles \cite{Ishikawa2005,Ishikawa2013,Ishikawa2014,Ishikawa2015,Ishikawa2016,Ishikawa2018,edery_wave_2021}. In this work, we will show that in spacetimes which are not translationally invariant there are always similar corrections.

Scattering calculations have been of recent interest in the literature \cite{Lankinen2017,Lankinen2018a,Lankinen2018b}, but many of these calculations have been done in translationally invariant spacetimes, where the issue investigated in this work does not appear -- though expanding universes have other well-known issues of their own\cite{Birrell1982}. Since S-matrix elements are fundamental components of any quantum field theory, we believe the corrections in spatially inhomogenous spacetimes are worth investigating.

We first examine the difference between the normalization methods in static spacetimes. We then provide an explicit example. Finally, we discuss implications.

\section{Packet and box normalizations}\label{sec:results}
We will now derive the curved spacetime box and wave packet normalization formulae for a single massive scalar particle of mass $M$ decaying to two scalars of some other -- possibly zero -- mass $m$ in a static spacetime $g$ in D+1 dimensions, $D \in \mathbb{N}$. We have chosen this particular process since it is the minimal example that illustrates our point, but of course similar reasoning applies to other processes in static spacetimes. We will perform the calculation at tree level, because the essential point does not change in higher orders and we wish to avoid dealing with renormalization. Since issues of normalizing the state are usually glossed over in flat spacetime quantum field theory, the reader may wish to consult appendix \ref{app:review} to see a simple 1+1 dimensional derivation of the usual decay rate formula in both the box and wave packet normalizations in full detail. Alternatively, those simple cases are found in less specific detail in Peskin\&Schroder \cite{Peskin1995} for wave packets or Folland \cite{Folland2008} for box normalization.

A few notes about decay rates are in order. In both the box and the wave packet normalization, we shall assume that the limit $T\rightarrow \infty$ makes sense. Unstable particles cannot really, of course, be created in the infinitely distant past. Moreover, the decay rate is usually defined in the reference frame where the momentum of the decaying particle is 0. Since we do not have a good definition of momentum in general, we cannot do this. The relation of $d\Gamma$ to flat spacetime decay rate is by analogy: we are really just dealing with an abstract S-matrix element. These issues, however, are of no concern to us here, as they are equally present in both the box and wave packet normalizations; for our purposes, the limit $T\rightarrow \infty$ is formal.

The D+1 dimensional Klein-Gordon equation for metric $g$ and mass $m$ (correspondingly $M$) is given in the minimal coupling by
\begin{align}
    \bigg( \frac{1}{\sqrt{-g}}\partial _\mu \bigg[ \sqrt{-g}g^{\mu \nu}\partial _\nu \bigg] +m^2\bigg) \phi (x) = 0. \label{eq:kg}
\end{align}
We could also use a more general equation, but this introduces nothing to our analysis. We will call the in-state field $\phi $ and the outgoing fields $\psi$ with associated annihilation (creation) operators $b_k (b_k^\dagger)$, $a_k (a_k^\dagger)$. The mode functions -- solutions of \eqref{eq:kg} -- are denoted $f^m_k (\mathbf{x})e^{iE_k(t)}$ for the out-state and $f^M_k(\mathbf{x})e^{i\omega _kt}$ for the in-state. This decomposition can always be found in static spacetimes \cite{Birrell1982}. The interaction is given by
\begin{align}
    \mathcal{L}_{\text{int}} = -\lambda \int d^{D+1}x \sqrt{-g(x)}\phi (x) \psi(x)^2.
\end{align}
where $-g(x)$ is the determinant of the metric. We emphasize that in a general static spacetime, the index $k$ of the mode functions is not a momentum -- it is just a variable indexing the states of the field. The full interpretation as momentum is only available in spatially flat spacetimes, where it is a conserved quantity. Our use of $k$, $p$, etc. for indexing is merely conventional \cite{unruh_notes_1976}. Our commutator is $[a _\mathbf{k}, a^\dagger_{\mathbf{k}'}] = (2\pi )^D2E_\mathbf{k}\delta^{(D)} (\mathbf{k}-\mathbf{k}')$.
\subsection{Wave packet normalization}\label{eq:wavepack}
Since we do not know the distribution of out-states, we will use wave packets that simulate a detector of finite resolution. We do this by smoothing the states over a D-dimensional cube of side length $\epsilon$. Hence our out-states are
\begin{align}
    |\Delta _{k}\rangle &= \int _{\Delta _k} \widehat{d\mathbf{p}} a^\dagger _\mathbf{p}|0\rangle ,\\
    \widehat{d\mathbf{p}}&=\frac{d^D\mathbf{p}}{\sqrt{(2\pi)^D 2E_\mathbf{p} V_\epsilon }}.
\end{align}
Here, $\Delta _k$ is the integral over a cube of volume $V_\epsilon$ centered on $\mathbf{k}$. This state is normalized:
\begin{align}
    \langle \Delta _k | \Delta _k\rangle = 1.
\end{align}
It also has the following useful Taylor expansion when $\epsilon\ll 1$:
\begin{align}
    \int _{\Delta _k} d\mathbf{p}h(\mathbf{p}) \approx V_\epsilon h(\mathbf{k}),\label{eq:outpacktrick}
\end{align}
where h($\mathbf{k}$) is a function smooth almost everywhere.

For the in-state, we will use a more general wave packet such that
\begin{align}
    |g^\sigma _\mathbf{k}\rangle &=  \int \widetilde{d\mathbf{p}}g^\sigma _\mathbf{k}(\mathbf{p}) a^\dagger _\mathbf{p}|0\rangle\\
    \widetilde{d\mathbf{p}}&=\frac{d^D\mathbf{p}}{\sqrt{(2\pi)^D 2E_\mathbf{p}}}
\end{align}
where we understand the integral to go over the whole k-spectrum. The central value of the wave packet is $\mathbf{k}$ and $\sigma$ its width. For normalization we require $\langle g_\mathbf{k}^\sigma | g_\mathbf{k}^\sigma\rangle = 1$, which leads to
\begin{align}
    \int d^D\mathbf{p}| g^\sigma_\mathbf{k}(\mathbf{p})|^2=1,
\end{align}
We demand that the wave packet is sharp. Practically, this means that
\begin{align}
    \int d\mathbf{p}g_\mathbf{k}^\sigma(\mathbf{p})h(\mathbf{p}) \approx h(\mathbf{k})\int d\mathbf{p}g_\mathbf{k}^\sigma(\mathbf{p}), \label{eq:sharp}
\end{align}
for a function $h$ smooth almost everywhere. This is a standard assumption found in e.g. \cite{Peskin1995}.
One wave packet satisfying the foregoing properties would be a Gaussian:
\begin{align}
    g^\sigma_\mathbf{k} (\mathbf{p}) = \frac{1}{(\pi \sigma)^{D/4}}\exp \bigg( \frac{(\mathbf{k}-\mathbf{p})^2}{2\sigma } \bigg).
\end{align}
With these preliminaries, we define the average decay rate as
\begin{align}
    \Gamma ^{p_1}_{k_1k_2} = \lim _{T\rightarrow \infty}\frac{1}{T}|\langle \text{out},\Delta _{k_1} \Delta _{k_2}|g^\sigma _\mathbf{p}, \text{in}\rangle|^2
\end{align}
where $T$ is the total coordinate time of the interaction. We can obtain the tree-level result for the amplitude by using, for example, the LSZ theorem in curved spacetime \cite{Birrell1979}:

\begin{align}
    &\langle \text{out},\Delta _{k_1} \Delta _{k_2}|g^\sigma _\mathbf{p}, \text{in}\rangle\\
    &= \int _{\Delta _{k_1}}\widehat{d\mathbf{k}}\int _{\Delta k_1}\widehat{d\mathbf{k}'}\int\widetilde{d\mathbf{p}}\int d^{D+1}\mathbf{x}_1d^{D+1}\mathbf{x}_2d\mathbf{y}f^{m}_{\mathbf{k}}(\mathbf{x}_1)f^{m}_{\mathbf{k}'}(\mathbf{x}_2)f^{M}_{\mathbf{k}'}(\mathbf{y})\mathcal{K}_{\mathbf{x}_1}^{m}\mathcal{K}_{\mathbf{x}_2}^{m}\mathcal{K}_{\mathbf{y}}^{M}\nonumber\\
    &\times \langle 0|T\phi (\mathbf{y})\psi (\mathbf{x}_1) \psi (\mathbf{x}_2)|0\rangle \sqrt{-g(x_1)}\sqrt{-g(x_2)}\sqrt{-g(y)}
\end{align}
with
\begin{align}
    \mathcal{K}_x^m &= \frac{1}{\sqrt{-g}}\partial _\mu \bigg[ \sqrt{-g}g^{\mu \nu}\partial _\nu \bigg] +m^2,\\
    \mathcal{K}_x^m G_F^m(x,y)&=\frac{\delta (x-y)}{\sqrt{-g(x)}},\\
    G_F^M(x,y) &= \langle 0|T\phi_0(x)\phi_0(y)|0\rangle ,
    \end{align}
where in the last equation $0$ indicates a free field and holds also for $\psi$ with the appropriate mass. The following analysis applies similarly to fermions and vectors, for which the formulae can be found in \cite{Huhtala_Vilja_2024}. At tree level, we see after an application of Wick's theorem that
\begin{align}
    &\Delta \Gamma ^{p_1}_{k_1k_2} = \lim _{T\rightarrow \infty}\frac{\lambda ^2}{T} \int_{\Delta _{k_1}} \widehat{d\mathbf{k}}\widehat{d\mathbf{k}'}\int _{\Delta _{k_2}} \widehat{d\mathbf{q}}\widehat{d\mathbf{q}'}\int \widetilde{d\mathbf{p}}\widetilde{d\mathbf{p}'}g_\mathbf{p_1}^\sigma(\mathbf{p})g_\mathbf{p_1}^\sigma (\mathbf{p}')\int d^{D+1}\mathbf{x}d^{D+1}\mathbf{x}' \nonumber\\
    &\times \bigg[4\sqrt{-g(x)}\sqrt{-g(x')} f^m_\mathbf{k}(\mathbf{x})f^m_{\mathbf{k}'}(\mathbf{x})f^m_\mathbf{q}(\mathbf{x})f^m_{\mathbf{q}'}(\mathbf{x'}) f^M_\mathbf{p}(\mathbf{x'})f^M_{\mathbf{p}'}(\mathbf{x'}) e^{i\Delta E t -i\Delta E' t'}\bigg].
\end{align}
with $\Delta E = E_k+E_{k'}-\omega _p$ and $\Delta E' = E_q+E_{q'}-\omega _{p'}$. We take $\Delta \Gamma ^{p_1}_{k_1k_2}$ to be the decay rate over the small $\mathbf{k}$ intervals; this will approach the differential decay rate as the intervals get smaller. We have chosen the spatial parts of the wave functions to be real with no loss of generality. We now use \eqref{eq:outpacktrick} and \eqref{eq:sharp} to obtain
\begin{align}
    \Delta \Gamma ^{p_1}_{k_1k_2} = &\frac{4\lambda^2V_{\epsilon_2}V_{\epsilon_1}}{(2\pi)^{3D}8E_{k_1}E_{k_2}\omega _p}\lim _{T\rightarrow \infty}\frac{1}{T} \bigg|\int _0^T e^{i(E_{k_1}+E_{k_2}-\omega _p) t} \bigg|^2\\
    &\times\bigg| \int d^D\mathbf{x}d^D\mathbf{x}' \sqrt{-g(x)}f^m_\mathbf{k_1}(\mathbf{x})f^m_{\mathbf{k_2}}(\mathbf{x})f^M_\mathbf{p_1}(\mathbf{x})\bigg|^2 \nonumber
   \int d\mathbf{p}d\mathbf{p}'g_\mathbf{p_1}^\sigma(\mathbf{p})g_\mathbf{p_1}^\sigma (\mathbf{p}').
\end{align}
The limit produces $2\pi \delta (E_{k_1}+E_{k_2}-\omega _{p_1})$, and thus we have
\begin{align}
    \Delta \Gamma ^{p_1}_{k_1k_2} = &\frac{4\lambda^2V_{\epsilon_1}V_{\epsilon_2}\delta (E_{k_1}+E_{k_2}-\omega _{p_1})}{(2\pi)^{3D-1}8E_{k_1}E_{k_2}\omega _p}\bigg|\int d\mathbf{x}d^D\mathbf{x}' \sqrt{-g(x)}f^m_\mathbf{k_1}(\mathbf{x})f^m_{\mathbf{k_2}}(\mathbf{x})f^M_\mathbf{p_1}(\mathbf{x})\bigg|^2\nonumber\\
   & \times\int d\mathbf{p}d\mathbf{p}'g_\mathbf{p_1}^\sigma(\mathbf{p})g_\mathbf{p_1}^\sigma (\mathbf{p}')\label{eq:wavepackresult}
\end{align}
As $V_{\epsilon_1} ,V_{\epsilon_2}$ get smaller, we obtain the differential decay rate in terms of $dk_1$ and $dk_2$.

The last term is a wave packet interference contribution which is not present in Minkowski spacetime. We are unable to rid ourselves of the wave packet dependence due to the lack of symmetries. In particular, it is the lack of translation invariance that leads to the extra term. Quantum field theories, and thus Fock space states, are defined on time-slices with equal-time commutators; the definition of the states themselves loses symmetry when the translation invariance is not present.

We now turn to box normalization.
\subsection{Box normalization}\label{sec:box}
The idea of normalizing states with a box is to discretize the spectrum of the Klein-Gordon operator on the left hand side of \eqref{eq:kg}. This is done by choosing some region of the manifold as the "box", limiting the theory to this box and then setting up boundary conditions that discretize the spectrum. We assume this is possible for static metrics of the kind we are interested in -- though it is not entirely obvious that every operator of the type \eqref{eq:kg} can be discretized this way. In this section, we will understand the mode functions $f_\mathbf{k}^m(\mathbf{x})$ to satisfy the appropriate boundary conditions. The appropriate commutation relation is now $[a _\mathbf{k}, a_{\mathbf{k}'}^\dagger] = V(2\pi)^D2E_\mathbf{k}\delta _{\mathbf{k}\mathbf{k}'}$ with $V$ the coordinate volume of the box.

Thus a normalized state is
\begin{align}
    |k_{\text{box}}\rangle = V^{-1/2}((2\pi)^{D}2E_{k})^{-1/2}a^\dagger _k |0\rangle .
\end{align}
Then, with the integral measure for the states being $(2\pi)^{D/2}Vd^D\mathbf{k}$ at the limit of large V, we have
\begin{align}
    d\Gamma ^{p_1}_{k_1k_2} = &\lim _{T\rightarrow \infty} \frac{1}{T}\frac{4V^2dk_1dk_2}{(2\pi)^{2D}8E_{k_1}E_{k_2}\omega_{p_1}V^3}\bigg| \int _0^T dt e^{i(E_{k_1}+E_{k_2}-\omega _{p_1})t} \bigg|^2\nonumber\\
    &\times \bigg| \int_{\text{box}}  d^D\mathbf{x} \sqrt{-g(x)}f^m_{\mathbf{k_1}}(\mathbf{x})f^m_{\mathbf{k_2}}(\mathbf{x})f^M_{\mathbf{p_1}}(\mathbf{x})\bigg|^2\nonumber\\
    &= \frac{dk_1dk_2 \delta (E_{\mathbf{k_1}}+E_{\mathbf{k_2}}-\omega _{\mathbf{p_1}})}{(2\pi)^{2D-1}2E_{k_1}E_{k_2}\omega_{p_1}V}\bigg| \int_{\text{box}}  d^D\mathbf{x} \sqrt{-g(x)}f^m_{\mathbf{k_1}}(\mathbf{x})f^m_{\mathbf{k_2}}(\mathbf{x})f^M_{\mathbf{p_1}}(\mathbf{x})\bigg|^2. \label{eq:boxnormresult}
\end{align}
In flat spacetimes, our choice of a box of constant $V$ would make no difference due to the translation invariance of the theory. In curved spacetimes, though, different regions of the theory are obviously not equivalent. The limit $V\rightarrow \infty$ is trivially easy to take in flat spacetime (it is of the same kind as the limit $T\rightarrow \infty$), but hard or impossible to see in general static spacetimes.

In flat spacetimes, wave packet and box normalizations produce the same end result, since the details of both choices vanish at the appropriate limits \cite{Folland2008, Peskin1995}. Clearly, \eqref{eq:boxnormresult} and \eqref{eq:wavepackresult} are not the same. The integral of mode functions is done over a limited subset of the spacetime, and the mode functions themselves are thus different (satisfying e.g. open boundary conditions in the box). The wave packet normalization, on the other hand, produces a wave-packet dependent interference term. Both of these fundamentally result from the lack of symmetry in a theory.

In box normalization, the end result of taking limit $V\rightarrow \infty$ is also sensitive to the global structure of the spacetime in the sense that to take the limit, the metric (and potential boundary conditions) have to be known in the entire spacetime. This is not true in the wave packet normalization, where the equivalent limit depends only on details of the wave packet, where we have considerable freedom. We will illustrate these points presently.

\section{Example}\label{sec:example}
In this section, we treat a simple model that illustrates the results of section \ref{sec:results}. We will use the 1+1 dimensional metric
\begin{align}
    g_{\mu\nu} = (A+B\tanh (x))\eta _{\mu\nu}\label{eq:fullexample}
\end{align}
with $A,B\in \mathbb{R}, A,B>0$ and $A+B\tanh (x)>0\ \forall x\in \mathbb{R}$. This metric is asymptotically Minkowskian as $x\rightarrow \infty, x\rightarrow -\infty$ and $L\in \mathbb{R},L>0$.
\subsection{Wave packet}
First, we need to solve \eqref{eq:kg} for metric \eqref{eq:fullexample}. This solution is already known when the metric depends on time instead of space, and exchanging time and space is trivial in 1+1 dimensional spacetime. Hence we only quote the result \cite{Birrell1982}:
\begin{align}
    f^{m,M}_{k}&(x)e^{i\omega _kt} \nonumber\\
    &= \mathcal{N}e^{i\omega _{k}t-ig_+x-ig_-}\ln (2\cosh (x)){}_{2}\text{F}_1\bigg(1+(ig_-),ig_-;1-ig_{\text{left}};\frac{1}{2}(1+\tanh (x)\bigg)
\end{align}
with
\begin{align}
    g_{\text{left}}&=(\omega _k^2+m^2(B-A))^{1/2},\\
    g_{\text{right}}&=(\omega _k^2-m^2(A+B))^{1/2},\\
    g_{\pm} &= \frac{1}{2}(g_{\text{left}}\pm  g_{\text{right}}),
\end{align}
and correspondingly with the incoming mass $M$. $\mathcal{N}$ is chosen so as to fulfill the quantization condition, and ${}_{2}\text{F}_1$ is a hypergeometric function. The solution is chosen so that asymptotically as $x\rightarrow -\infty$, it reduces to the flat spacetime plane wave solution.  There is another hypergeometric solution with similar parameters that reduces to the a plane wave in $x\rightarrow \infty$, but we will not explicitly write it here. Note that the energy $\omega _k$ depends on $k$, the variable which indexes our states, in some complicated way that does not concern us here; in this case, the variable is continuous.

We are faced with the choice of wave packet. There is really not much reason to choose one over the other; if momentum were a good variable, we might think that a Gaussian, normal-distributed set of initial momenta is physically reasonable. Alas, the space has no translation invariance, and so this logic does not straightforwardly work. 

We do, however, have an asymptotically flat spacetime, so we might guess that a Gaussian is a reasonable choice anyway, at least outside of the immediate vicinity of $x=0$. In that case, we get from \eqref{eq:wavepackresult}
\begin{align}
    \Delta \Gamma ^{p_1}_{k_1k_2} = &\frac{4\lambda^2V_{\epsilon_1}V_{\epsilon_2}\delta (E_{k_1}+E_{k_2}-\omega _{p_1})}{(2\pi)^{2}8E_{k_1}E_{k_2}\omega _p}\bigg|\int dxdx' (A+B\tanh (x))f^m_\mathbf{k_1}(\mathbf{x})f^m_{\mathbf{k_2}}(\mathbf{x})f^M_\mathbf{p_1}(\mathbf{x})\bigg|^2\nonumber\\
   & \times\int dpdp'\frac{1}{(\pi \sigma)^{1/2}}\exp \bigg(-\frac{(p-p_1)^2}{2\sigma }\bigg)\exp \bigg (\frac{-(p'-p_1)^2}{2\sigma}\bigg).
\end{align}
We have not written out the hypergeometric functions explicitly, as there is no hope of analytically computing the integral. The overlap term is
\begin{align}
    \int dpdp'\frac{1}{(\pi \sigma)^{1/2}}\exp \bigg(-\frac{(p-p_1)^2}{2\sigma }\bigg)\exp \bigg (\frac{-(p'-p_1)^2}{2\sigma}\bigg) =\sqrt{\pi \sigma}.
\end{align}
Hence, we are left with an explicit wave packet dependence. In other words, we must now define our decay rate with respect to a particular choice of wave packet.
\subsection{Box normalization}
We note immediately that the choice of box makes a crucial difference. Suppose we take the box $[x_0-\frac{1}{2}L,x_0+\frac{1}{2}L]$ with $x_0\in \mathbb{R},x_0 \gg 1$. Then
\begin{align}
    g_{\mu\nu}(x) = (A+B\tanh(x))\eta _{\mu\nu} \approx (A+B)\eta _{\mu\nu},\label{eq:exmetric}
\end{align}
inside the box. Thus, in this box, we have in \eqref{eq:boxnormresult} with periodic boundary conditions
\begin{align}
    \frac{1}{V}\bigg| \int_{\text{box}}  d^D\mathbf{x} \sqrt{-g(x)}f^m_{\mathbf{k_1}}(\mathbf{x})f^m_{\mathbf{k_2}}(\mathbf{x})f^M_{\mathbf{p_1}}(\mathbf{x})\bigg|^2=\frac{1}{V}\bigg| \int_{\text{box}}d\mathbf{x}(A+B)e^{i(k_1+k_2-ip_1)x}\bigg|^2 
\end{align}
with $k_i,p_i$ discretized with e.g. $k_1=n\pi /L$. This is just the flat spacetime result with a scaling factor $A+B$. On the other hand, suppose that $x_0 \approx 0$. Then a reasonable approximation is
\begin{align}
    g_{\mu\nu}(x) \approx A+Bx
\end{align}
and the solution to \eqref{eq:kg} (with open boundary conditions where $f^m_i (-L/2)=f^m_i(L/2)=0$) is
\begin{align}
    f^{m}_{i}(x) &= \mathcal{N}\bigg[ \text{Ai}(y(x,\omega _i))\text{Bi}(y(-L,\omega _i))-\text{Bi}(y(,\omega _i))\text{Ai}(y(-L,\omega _i))\bigg],\\
    y(x,\omega) &= \frac{Am^2-\omega _i^2+Bm^2x}{(bm^2)^{2/3}} .
\end{align}
Here $\mathcal{N}$ is a normalization factor chosen so that the commutation relations of our QFT are satisfied. Such a $\mathcal{N}$ can always be found since the equation \eqref{eq:kg} with metric \eqref{eq:exmetric} is a regular Sturm-Liouville problem, and therefore the solutions form an orthogonal set. There would then be a corresponding solution for the mass $M$.  Note that $\omega _i$ -- the set of allowed energies -- is now discrete and has to be chosen so as to satisfy the boundary conditions.

Thus, different choices of box (and boundary conditions!) produce rather different results. This would not be a problem if we could just take the solution for the whole space, restrict it to a box and ultimately take the limit once the integrals have been computed. This procedure works in flat spacetime, but is unfortunately far from trivial in general static spacetime. How would one go about taking the limit $V\rightarrow \infty$ for complicated integrals over hypergeometric functions, for instance? In practical terms, the best one can do is to choose a large box and compute the ingterals in it, but the choices of box then define differing decay rates.
\section{Discussion}\label{sec:discussion}
We have defined the decay rate of an unstable particle in two different ways found in the flat spacetime literature and found that these methods are not equivalent in curved spacetimes. We have also demonstrated this effect with a concrete example.

We emphasize that these results are really about the formal properties of scattering matrix elements in a curved spacetime QFT. Interpreting quantities like our $\Gamma ^{p_1}_{k_1k_2}$ as decay rates is tricky in curved spacetimes, because we cannot be sure that the limit $T\rightarrow \infty$ makes sense in an arbitrary spacetime. The states are moreover not labeled by momenta. All of these issues are caused by the lack of symmetry inherent in curved space QFTs, and each of them has to be investigated on a case-by-case basis once a spacetime has been chosen. For instance, the metric \eqref{eq:exmetric} provides at least asymptotic ($x\ll 1, x\gg 1$) definitions of momentum.

There is one more notable way of normalizing the Fock space states, and that is discretizing the spacetime, i.e. putting it on a lattice. However, there is no way to discretize a manifold while preserving all of its geometric properties (see \cite{crane_glimpse_2017} and references therein); some information about the curvature is inevitably lost in the process of discretization, and thus the limit of zero lattice spacing does not necessarily recover the correct continuous manifold.

We computed the decay rate to the first order tree-level only. Of course, the same analysis would apply for higher orders with loops, except the procedure would now be technically complicated. It seems inevitable that in higher orders e.g. trying to take the limit $V\rightarrow \infty$ would be even more challenging than at tree-level, but this would not fundamentally change our result.

Another question we did not address is the validity of the sharp wave packet assumption, eq. \eqref{eq:sharp}. In principle, we are free to choose any wave packet that satisfies the normalization conditions and possible boundary conditions, and thus one might imagine we can typically choose a wave packet that satisfies \eqref{eq:sharp}. However, in flat spacetime the wave packet would determine the initial distribution of momenta, so not every wave packet is equally as physically reasonable. A normal-distributed set of states seems plausible, but a Gaussian might not be a valid choice in every spacetime due to boundary conditions, nor is it a priori an appropriate initial distribution even in the ones where it is technically allowed. We also note that when defining quantum fields formally as operator-valued distributions, one uses classes of test functions; in the general curved spacetime case, these must be functions of compact support \cite{Wald1994}, thus ruling out e.g. Gaussian packets. If we cannot assume \eqref{eq:sharp}, then the result is instead
\begin{align}
    &\Gamma ^{p_1}_{k_1k_2} = \lim _{T\rightarrow \infty}\frac{4V^{k_1}_\epsilon V^{k_2}_\epsilon\lambda ^2}{T} \int \widetilde{d\mathbf{p}}\widetilde{d\mathbf{p}'}g_\mathbf{p_1}^\sigma(\mathbf{p})g_\mathbf{p_1}^\sigma (\mathbf{p}')\int d^{D+1}\mathbf{x}d^{D+1}\mathbf{x}' \sqrt{-g(x)}\sqrt{-g(x')}\nonumber\\
    &\times \bigg[ f^m_\mathbf{k}(\mathbf{x})f^m_{\mathbf{k}'}(\mathbf{x})f^m_\mathbf{q}(\mathbf{x})f^m_{\mathbf{q}'}(\mathbf{x'})f^M_\mathbf{p}(\mathbf{x'})f^M_{\mathbf{p}'}(\mathbf{x'}) e^{i\Delta E t -i\Delta E' t'}\bigg],
\end{align}
which is considerably more complicated. Note that even the limit $T\rightarrow \infty$ is now not easy to take, since the exponential time parts have different energy arguments.

These results have implications for any scattering calculation done in spacetimes in which momentum is not conserved and the states are labeled by a continuous variable (and thus in need of normalizing). Clearly, the foregoing reasoning applies just as well to e.g. the scattering of two particles or other multi-particle processes. The result is, then, of general formal importance: if we wish to be careful in defining our Fock states, we must inevitably make a choice of how to normalize them, and these choices are not equivalent. Hence, scattering elements are only defined with respect to a particular normalization scheme. The work of Ishikawa et al \cite{Ishikawa2005,Ishikawa2013,Ishikawa2018} and Edery \cite{edery_wave_2021} shows that wave packets cause extra terms to appear even in flat spacetimes under certain conditions; we have showed that extra terms of a similar type in fact appear always when momentum is not conserved.

\bibliography{sn-bibliography}
\appendix
\section{Flat spacetime formulae}\label{app:review}
We present here 1+1 dimensional derivations for flat spacetime decay rate formulae. This is for the reader's convenience, as all the details are usually not spelled out in textbooks, since in flat spacetimes it does not matter which way you normalize your states.
\subsection{Wave packets}
Using the conventions of section \ref{sec:results} explicitly with a Gaussian, we get 
\begin{align}
    &\Gamma ^{p_1}_{k_1k_2} = \lim _{T\rightarrow \infty}\frac{\lambda ^2}{T} \int_{\Delta _{k_1}} \widehat{dk}\widehat{dk'}\int _{\Delta _{k_2}} \widehat{dq}\widehat{dq'}\int \widetilde{dp}\widetilde{dp'}g_{p_1}^\sigma(p)g_{p_1}^\sigma (p')\nonumber\\
    &\times \bigg[4\int d^2xd^2x' e^{ikx-iE _k t}e^{ik'x-iE _{k'} t}e^{iqx-iE_q t}e^{iq'x-iE_{q'} t}e^{-ipx+i\omega _pt} e^{-ip'x+i\omega _{p'}t}\bigg]\\
    &\approx \frac{4V^{k_1}_\epsilon V^{k_2}_\epsilon}{(2\pi)^24E_{k_1}E_{k_2}}  \lim _{T\rightarrow \infty}\frac{\lambda ^2}{T}\int \widetilde{dp}\widetilde{dp'}g_{p_1}^\sigma(p)g_{p_1}^\sigma (p')(2\pi)^2\delta (k_1+k_2-p)\delta (k_1+k_2-p')\nonumber\\
    &\times \bigg[\int_0^T dt dt' e^{i(E_{k_1}+E_{k_2}-\omega _p)t-i(E_{k_1}+E_{k_2}-\omega _{p'})t}\bigg].
\end{align}
In the second line we used $\eqref{eq:outpacktrick}$ and integrated over the spatial coordinate. Assuming now $\eqref{eq:sharp}$ we find
\begin{align}
    \Gamma ^{p_1}_{k_1k_2}&=\frac{4V^{k_1}_\epsilon V^{k_2}_\epsilon}{(2\pi) 8E_{k_1}E_{k_2}\omega _{p_1}}  \lim _{T\rightarrow \infty}\frac{\lambda ^2}{T}\int dpdp'g_{p_1}^\sigma(p)g_{p_1}^\sigma (p')\delta (k_1+k_2-p)\delta (k_1+k_2-p')\\
    &\times \bigg|\int_0^T dt  e^{i(E_{k_1}+E_{k_2}-\omega _{p_1})t}\bigg|^2.
\end{align}
Performing the integrals over $p,p'$ and taking the limit, we find
\begin{align}
    \Gamma ^{p_1}_{k_1k_2}&=\frac{4V^{k_1}_\epsilon V^{k_2}_\epsilon\lambda ^2}{8E_{k_1}E_{k_2}\omega _{p_1}} |g^\sigma _{p_1}(k_1+k_2)|^2\delta (E_{k_1}+E_{k_2}-\omega _{p_1}).
\end{align}
As we approach the limit of sharp in and out states with $\epsilon \rightarrow 0$ and $\sigma \rightarrow 0$, we find
\begin{align}
    \Gamma ^{p_1}_{k_1k_2}&=\frac{dk_1dk_2\lambda ^2}{2E_{k_1}E_{k_2}\omega _{p_1}}\delta^{(2)} (p_1-k_1-k_2)
\end{align}
where we have used the fact that the limit of a Gaussian as its width approaches 0 is a Dirac delta function. This is the standard result to tree level in the interaction we used.
\subsection{Box normalization}
We use the same conventions as in section \ref{sec:box}. Specializing to 1+1, we have
\begin{align}
    d\Gamma ^{p_1}_{k_1k_2}=\frac{dk_1dk_2 \delta (E_{\mathbf{k_1}}+E_{\mathbf{k_2}}-\omega _{\mathbf{p_1}})}{(2\pi)2E_{k_1}E_{k_2}\omega_{p_1}V}\bigg| \int_{0}^V  d\mathbf{x} f^m_{\mathbf{k_1}}(\mathbf{x})f^m_{\mathbf{k_2}}(\mathbf{x})f^M_{\mathbf{p_1}}(\mathbf{x})\bigg|^2 
\end{align}
We have chosen the box to go from 0 to $V$; this choice is arbitrary as the mode functions are translation invariant. Since we are now in flat spacetime, we know the spatial part of these functions explicity; they are of course
\begin{align}
    f^{m,M}_{k}(x)=\exp (ikx)\ \ \text{with}\ \  k = \frac{n\pi}{L},\quad n \in \mathbb{Z}.
\end{align}
The limit of sharp momentum states is equivalent to taking $V\rightarrow \infty$. Hence, we compute:
\begin{align}
    \lim_{V\rightarrow \infty}\frac{1}{V}\bigg| \int_{0}^V  d\mathbf{x} f^m_{\mathbf{k_1}}(\mathbf{x})f^m_{\mathbf{k_2}}(\mathbf{x})f^M_{\mathbf{p_1}}(\mathbf{x})  \bigg|^2 = 2\pi \delta (k_1+k_2-p_1)
\end{align}
Thus we end up with the result
\begin{align}
    \Gamma ^{p_1}_{k_1k_2} = \frac{dk_1dk_2\lambda ^2}{2E_{k_1}E_{k_2}\omega _{p_1}}\delta^{(2)} (p_1-k_1-k_2),
\end{align}
which is the also the wave packet result, as expected.
\end{document}